\documentclass[%
 aip,
 apl,
 amsmath,amssymb,
 reprint,%
]{revtex4-1}

\usepackage{graphicx}
\usepackage{dcolumn}
\usepackage{bm}

\usepackage[utf8]{inputenc}
\usepackage[T1]{fontenc}
\usepackage{mathptmx}
\usepackage{hyperref}
\hypersetup{
    colorlinks=true,
    linkcolor=blue,
    filecolor=blue,      
    urlcolor=blue,
    citecolor=blue,
}
 
\newcommand{\units}[1]{\ensuremath{\,\mathrm{#1}}}
\begin{document}

\preprint{}

\title[]{Spin-valve Josephson junctions with perpendicular magnetic anisotropy for cryogenic memory}

\author{N~Satchell}
\affiliation{School of Physics and Astronomy, University of Leeds, Leeds, LS2 9JT, United Kingdom}

\author{PM~Shepley}
\affiliation{School of Physics and Astronomy, University of Leeds, Leeds, LS2 9JT, United Kingdom}

\author{M~Algarni}
\affiliation{School of Physics and Astronomy, University of Leeds, Leeds, LS2 9JT, United Kingdom}

\author{M~Vaughan}
\affiliation{School of Physics and Astronomy, University of Leeds, Leeds, LS2 9JT, United Kingdom}

\author{E~Darwin}
\affiliation{School of Physics and Astronomy, University of Leeds, Leeds, LS2 9JT, United Kingdom}

\author{M~Ali}
\affiliation{School of Physics and Astronomy, University of Leeds, Leeds, LS2 9JT, United Kingdom}

\author{MC~Rosamond}
\affiliation{School of Electronic and Electrical Engineering, University of Leeds, Leeds LS2 9JT, United Kingdom}

\author{L~Chen}
\affiliation{School of Electronic and Electrical Engineering, University of Leeds, Leeds LS2 9JT, United Kingdom}

\author{EH~Linfield}
\affiliation{School of Electronic and Electrical Engineering, University of Leeds, Leeds LS2 9JT, United Kingdom}

\author{BJ~Hickey}
\affiliation{School of Physics and Astronomy, University of Leeds, Leeds, LS2 9JT, United Kingdom}

\author{G~Burnell}
\email{g.burnell@leeds.ac.uk}
\affiliation{School of Physics and Astronomy, University of Leeds, Leeds, LS2 9JT, United Kingdom}

\date{\today}

\begin{abstract}

We demonstrate a Josephson junction with a weak link containing two ferromagnets, with perpendicular magnetic anisotropy and independent switching fields in which the critical current can be set by the mutual orientation of the two layers. Such pseudospin-valve Josephson junctions are a candidate cryogenic memory in an all superconducting computational scheme. Here, we use Pt/Co/Pt/CoB/Pt as the weak link of the junction with $d_\text{Co} = 0.6\units{nm}$, $d_\text{CoB} = 0.3\units{nm}$, and $d_\text{Pt} = 5\units{nm}$ and obtain a $60\%$ change in the critical current for the two magnetization configurations of the pseudospin-valve.  Ferromagnets with perpendicular magnetic anisotropy have advantages over magnetization in-plane systems which have been exclusively considered to this point, as in principle the magnetization and magnetic switching of layers in the junction should not affect the in-plane magnetic flux.

\end{abstract}

\maketitle

Josephson junctions containing ferromagnetic weak links have been of interest over the last twenty years due to the additional physics present when pair correlations from the superconductor (\textit{S}) interact with the exchange field of the ferromagnet (\textit{F}) \cite{RevModPhys.77.935,RevModPhys.77.1321,eschrig_spin-polarized_2011,linder_superconducting_2015,0034-4885-78-10-104501}. Examples include the tuning of the ground state phase difference across a junction from 0 to $\pi$ by changing the thickness of the \textit{F} layer \cite{buzdin1982critical,PhysRevLett.86.2427,PhysRevLett.89.137007,PhysRevLett.97.177003}. The additional physics can also drive the generation of $m_s =\pm 1$ spin-triplet pair correlations with spin projection along the magnetization axis of the \textit{F} layer in the junction \cite{PhysRevLett.86.4096}, leading to pair propagation through the \textit{F} layer over much longer distances than the singlet component \cite{keizer2006spin,PhysRevLett.104.137002,robinson2010controlled,PhysRevB.82.100501,PhysRevLett.116.077001,lahabi2017controlling,doi:10.1098/rsta.2015.0150}.

By adding a second \textit{F} layer, $S-F_1-F_2-S$ pseudospin-valve (PSV) Josephson junctions offer functional devices where the ground state phase difference and critical current across the junction can be tuned by controlling the mutual orientations of the \textit{F} layers \cite{PhysRevLett.86.3140,PhysRevB.64.172511,Golubov2002,PhysRevB.66.140503,PhysRevB.69.024525,PhysRevB.74.184509,PhysRevB.75.054503}. One application of these devices is for cryogenic memory, which is needed both to improve energy efficiency of large-scale computation and as part of the effort to interconnect classical computers with quantum computers \cite{6449287,soloviev2017beyond,mcdermott2018quantum}. The first experimental report by Bell \textit{et al.} used a Co/Cu/Permalloy PSV to control the critical current of the junction \cite{bell_controllable_2004}. Promising electrical engineering architectures which integrate the ferromagnetic Josephson junction bits into scalable memory cells compatible with current technology exists, and recently PSV devices have been demonstrated by several groups \cite{1439786,baek2014hybrid, doi:10.1063/1.4862195,PhysRevApplied.3.011001,  gingrich_controllable_2016,dayton2017experimental, PhysRevB.97.024517,Glickeaat9457, Madden_2018,10.1117/12.2321109}. To date, all experimental works on PSV Josephson junctions have considered in-plane \textit{F} layers as used by Bell \textit{et al.}.

Here, we construct devices where both \textit{F} layers have perpendicular magnetic anisotropy (PMA). In order to achieve independent switching of the two \textit{F} layers, we use the standard approach of having two different materials as `hard' and `soft' magnetic layers, here CoB and Co. 
Whilst Co has been widely studied in Josephson junctions\cite{PhysRevLett.97.177003,PhysRevB.76.094522,PhysRevB.80.020506,satchellSOC2018}, the amorphous alloy CoB has not previously been used. Both Co/Pt and CoB/Pt interfaces exhibit PMA giving an overall PMA for thin ferromagnetic layers. Previous work using Pt as an interlayer in Co/Ni Josephson junctions found the optimal Josephson current was achieved for $d_\text{Pt} \geq 4.5\units{nm}$ \cite{PhysRevB.99.174519}. Here we use $d_{Pt} = 5\units{nm}$ which is adequate to buffer the fcc growth of the Co, generate the PMA and have the Pt act as a spacer to separate the magnetic switching of the $F$ layers. 

Samples were fabricated into standard ``sandwich'' planar Josephson junctions using a three stage photolithographic method. First, we defined the bottom electrode stencil in a LOR 7B/S1813 bilayer resist. The [Nb/Au]$_\text{x3}$/Nb bottom electrode, Pt/Co/Pt/CoB/Pt weak link and Nb/Au cap were deposited by dc magnetron sputtering in the Royce Deposition System \cite{Royce}. The magnetrons were mounted below, and confocal to, the substrate with source-substrate distances of 134~mm. The base pressure of the vacuum chamber was 1$\times$10$^{-9}\units{mBar}$. The samples were deposited at room temperature with an Ar (6N purity) gas pressure of 3.6$\times$10$^{-3}\units{mBar}$ for the Nb and Au layers and 4.8$\times$10$^{-3}\units{mBar}$ for the weak link. The [Nb/Au]$_\text{x3}$/Nb superlattice was used for the base electrode as the superlattice has a lower surface roughness compared to a single Nb layer of comparable total thickness \cite{PhysRevB.85.214522,PhysRevB.99.174519,Quarterman}.

In the second step, a circular S1813 resist mesa of diameter either 3 or $4\units{\mu m}$ was defined as a mask for broad beam $\mathrm{Ar^+}$ ion milling, which removed the cap and layers comprising the weak link from the bottom electrode except in the area of the junction. The size of the resist mesa mask defines the junction area. This was followed by deposition of $50\units{nm}$ SiO$_\text{x}$ insulator by RF sputtering to isolate the bottom electrode.  

In the third and final step, the top electrode stencil was defined in S1813 resist, and the top electrode, 150 nm of Nb, was sputter deposited immediately following an \textit{in-situ} $\mathrm{Ar^+}$ ion mill to ensure a clean interface between the junction mesa and top electrode. The full structure of the final device with thickness in (nm) was [Nb (25)/Au(2.5)]$_\text{x3}$/Nb (20)/Pt (5)/Co (0.6)/Pt (5)/CoB (0.3)/ Pt (5)/Nb (5)/Au (5)/ Nb (150). 

After fabrication, devices were measured in a continuous flow $^4\mathrm{He}$ cryostat with 3\units{T} horizontal superconducting Helmholtz coils. The sample can be rotated between in-plane and out-of-plane applied field. Traditional 4-point-probe transport geometry was used to measure the current-voltage characteristic of the junction with combined Keithley 6221-2182A current source and nano-voltmeter. Magnetization loops of sheet films were measured using a Quantum Design MPMS 3 SQUID magnetometer.

\begin{figure}
\includegraphics[width=0.9\columnwidth]{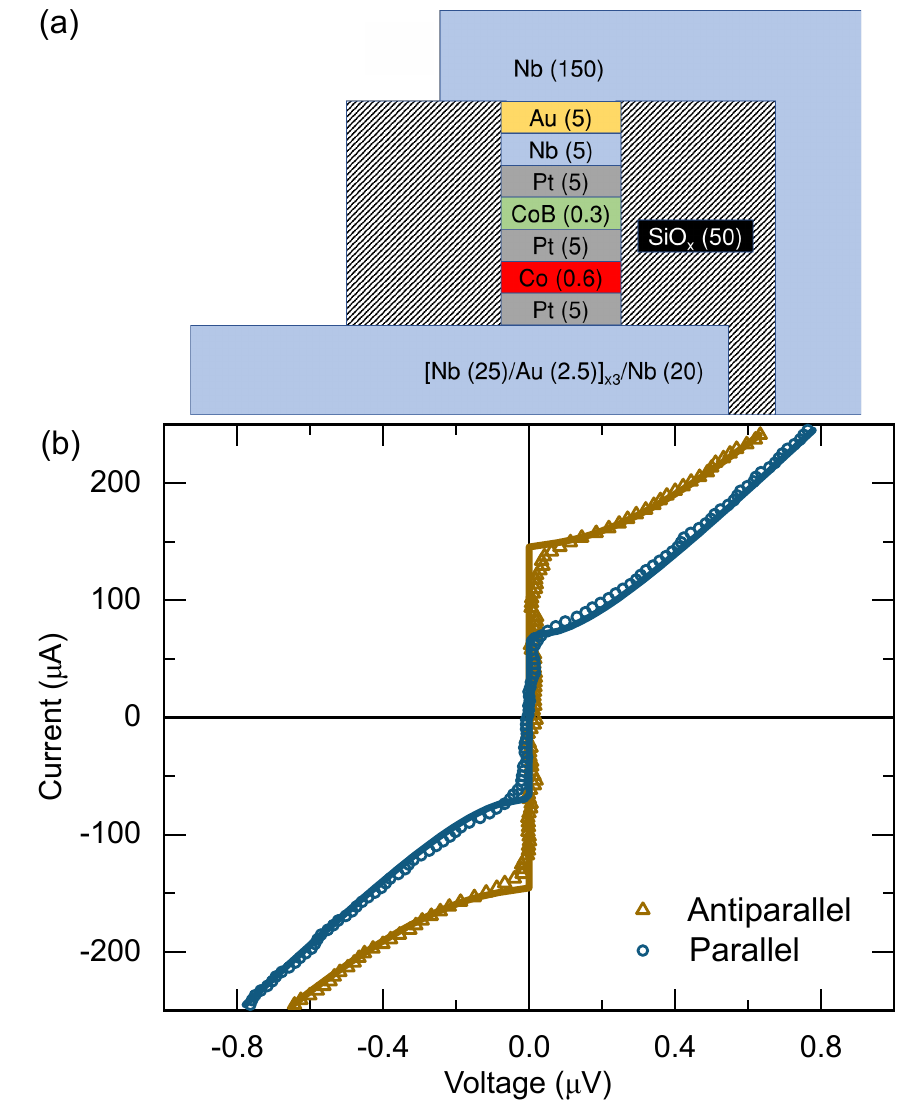}
\caption{\label{1} (a) Schematic cross section of the pseudospin-valve Josephson junction device, the thickness of each layer is given in nm (not to scale). (b) Current-voltage characteristic of the device measured at 0 applied field at 1.5 K after an applied field history at 15 K to set the device in either the parallel or antiparallel magnetic configurations.}
\end{figure}

Figure \ref{1} (a) shows the schematic of our final device and the full structure of our $S-F_1-F_2-S$ multilayer. The Au layers in the structure were sufficiently thin to be heavily proximitized by the adjacent Nb layers. The superconducting electrodes on either side of the weak link were split into separate voltage and current lines. Exemplar \textit{I-V} curves are shown in Figure \ref{1} (b), which demonstrates the essential function of our device. Measured at a temperature of 1.5 K at zero applied field, the critical current of the junction can be tuned by the relative orientation of the two \textit{F} layers. The \textit{I-V} characteristics of our devices follow the standard square-root form, $V=R_N\sqrt{I^2 - I^2_{c}}$, for $I \geq I_{c}$ as expected for over-damped Josephson junctions\cite{barone1982physics} and fits to the data are shown by solid lines in the figure.

\begin{figure}
\includegraphics[width=0.9\columnwidth]{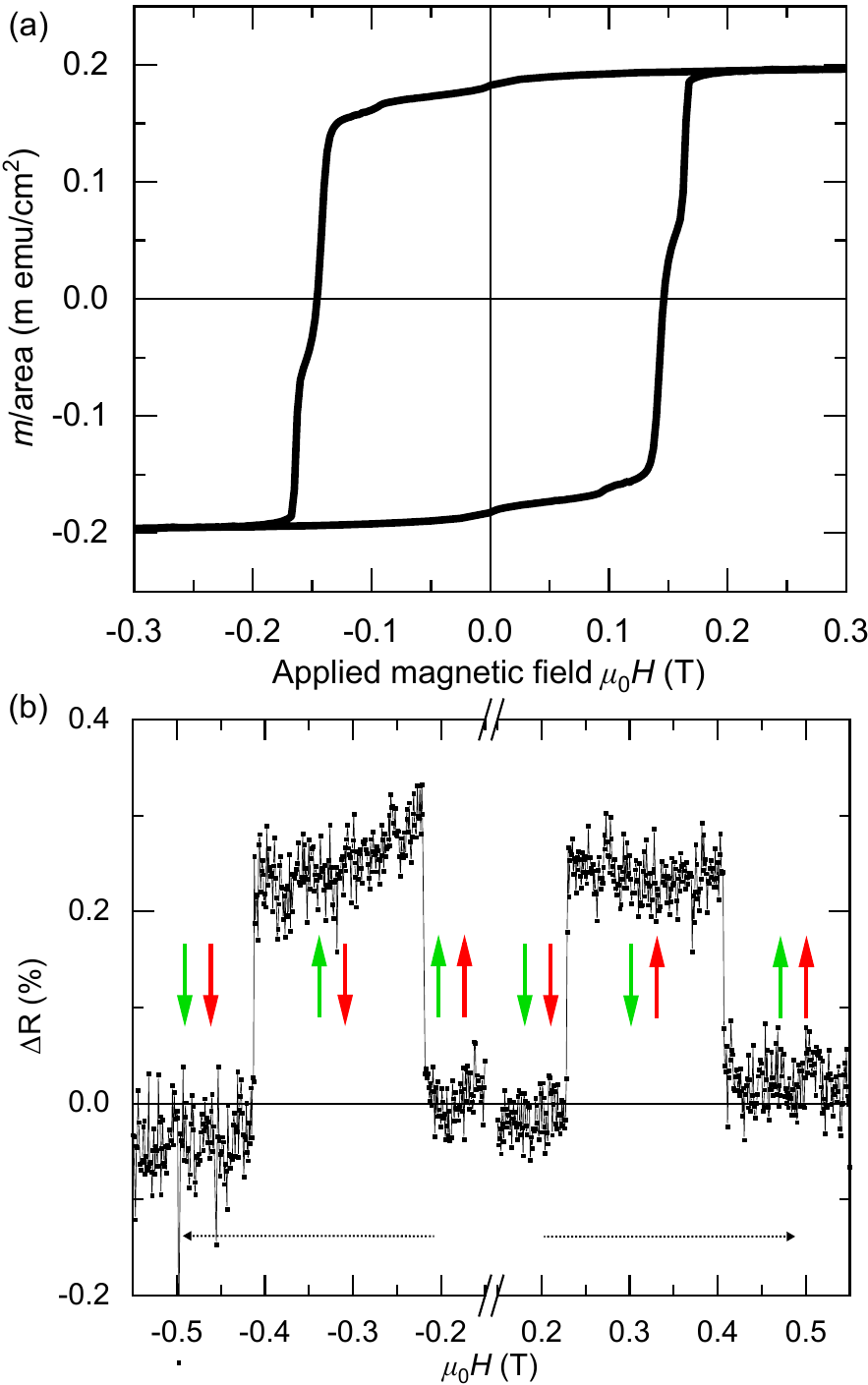}
\caption{\label{2} Magnetic switching behaviour of our structure. (a) The out-of-plane magnetic hysteresis of sheet films at 10\units{K}, showing a plateau at about $(-)0.15\units{T}$ where only one layer has switched. (b) Electrical transport above the critical current of our device at 1.5\units{K}. $\Delta$R is the difference in junction resistance between the increasing and decreasing magnetic field sweeps. The high resistance state corresponds to the layers being aligned antiparallel (indicated by the solid arrows). The dashed arrows indicate the direction of applied field sweep.}
\end{figure}

We next report the out-of-plane magnetic switching behaviour of our devices, which must be known and well controlled in order to achieve the two states shown in Figure \ref{1} (b). Figure \ref{2} (a)  shows the switching properties of a sheet film sample. The two main magnetic switching events occur close to applied field of $(-)0.15\units{T}$. First the Co layer, which has the larger magnetic moment ($m$), switches. Then there is a small plateau in $m$/area where the two layers align antiparallel before the second, thinner, CoB layer switches with the applied field \cite{Kikuchi2011,Metaxas2010,Metaxas2009,Moritz2004}. The small drop in $m$/area close to zero applied field is an artifact from our SQUID and is visible in these data due to the small signals from the thin magnetic layers under study. Dividing the $m$/area by the nominal total thickness of the \textit{F} layers (0.9 nm) gives a saturation magnetization of about 2190~emu/cm$^3$, which is significantly higher than the saturation magnetization of either bulk Co or CoB. This is due to the additional contribution from polarized Pt in proximity to the \textit{F} layers \cite{PhysRevB.65.020405}.

\begin{figure*}
\includegraphics[width=1.75\columnwidth]{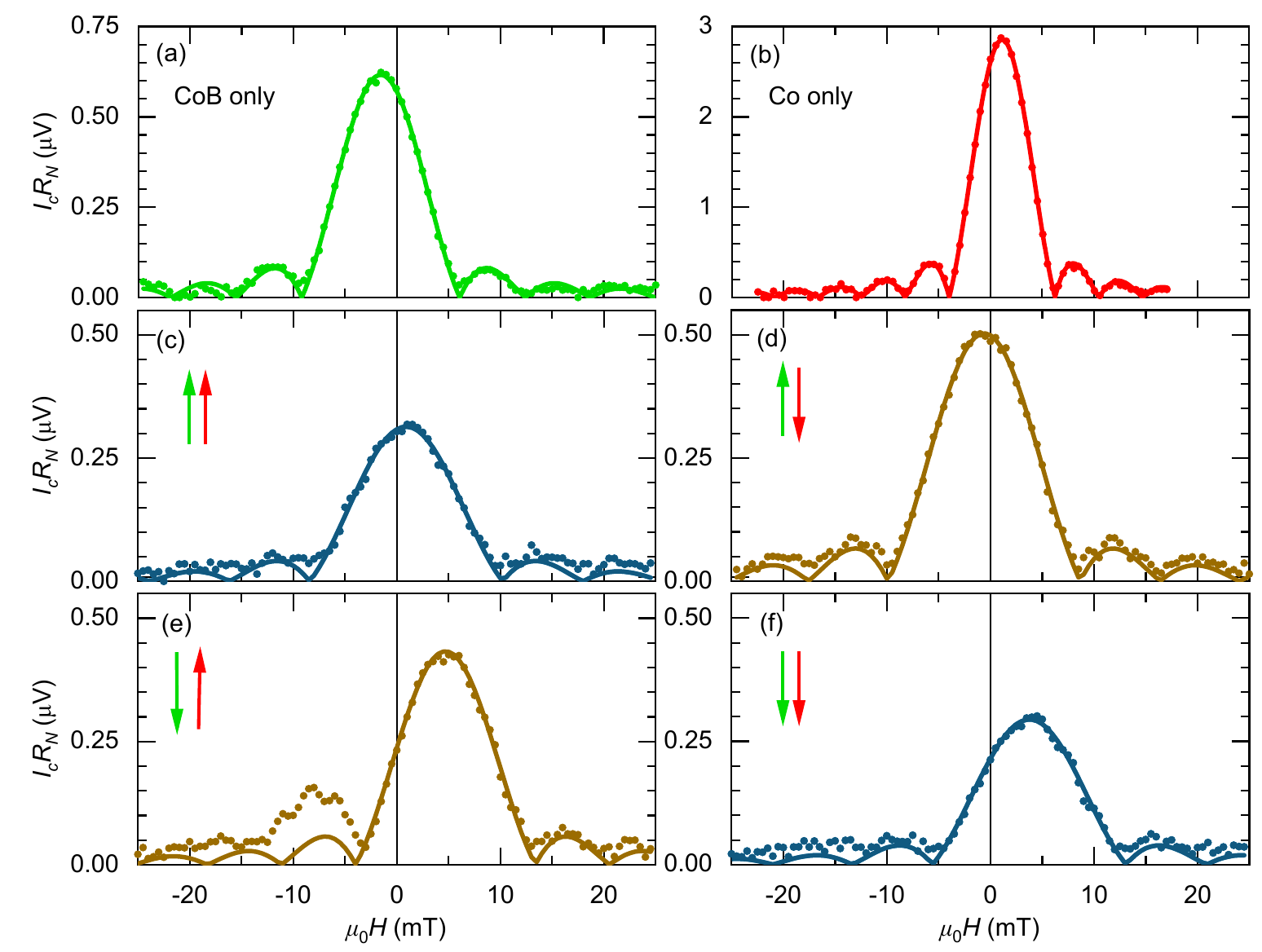}
\caption{\label{3}Product of the critical current and normal state resistance $I_cR_N$ vs field (the Fraunhofer pattern) measured at 1.5~K for (a,b) single ferromagnetic layer junctions containing (a) Pt (10\units{nm})/CoB (0.3\units{nm})/Pt (5\units{nm}) and (b) Pt (5\units{nm})/Co (0.6\units{nm})/Pt (10\units{nm}) weak links. (c-f) The pseudospin-valve device (shown in the schematic of Figure \ref{1} (a)) after field history to set up (c,f) the parallel magnetic configurations and (d,e) the antiparallel magnetic configurations. The field history applied to set up the magnetic configurations is described in the text.}
\end{figure*}

\begin{table*}[]
\begin{tabular}{l|l|l|l|l}
Sample (state) & $I_{c}R_N$ ($\mu$V)& $AR_N$ (f$\Omega$m$^2$) & $w$ ($\mu$m) & $J_c$ (MAm$^{-2}$) \\ \hline
PSV ($\uparrow \uparrow$)  & $0.312\pm0.001$  &  $9.45\pm0.03$  & $1.955\pm0.003$  & $33.1\pm0.1$   \\ \hline
PSV ($\uparrow \downarrow$)   &  $0.504\pm0.001$ &  $10.01\pm0.03$  & $2.005\pm0.002$  &  $50.4\pm0.1$  \\ \hline
PSV ($\downarrow \uparrow$)   & $0.439\pm0.001$  & $10.40\pm0.03$   & $2.046\pm0.002$  &  $42.2\pm0.1$  \\ \hline
PSV ($\downarrow \downarrow$) & $0.296\pm0.001$ &  $9.82\pm0.04$   & $1.995\pm0.003$  & $30.2\pm0.1$   \\ \hline
Pt-CoB-Pt      & $0.622\pm0.001$  &  $13.19\pm0.03$  & $2.441\pm0.002$  & $47.1\pm0.1$   \\ \hline
Pt-Co-Pt       & $2.864\pm0.002$  &  $14.59\pm0.03$  & $3.600\pm0.002$  & $196.3\pm0.3$   \\ 
\end{tabular}
\caption{\label{table1} Experimentally determined values corresponding to the data in Figure \ref{3} and fits to Equations \ref{Airy} and \ref{Phi}. $I_cR_N$ is the product of the maximum critical current and normal state resistance. $AR_N$ is the product of the area and normal state resistance. $w$ is the fitted width of the junction. $J_c$ is the maximum critical current density. Reported errors are statistical fitting uncertainties.}
\end{table*}
 
We perform electrical detection of the magnetic switching in the device structure at the temperature of interest (1.5\units{K}), Figure \ref{2} (b). Our measurement geometry uses the standard approach for current perpendicular-to-plane magnetoresistance (CPP-MR). An overview of the CPP-MR method, including shortcomings, are given in the review articles \cite{BASS1999274,bass2007spin,BASS2016244}. By using an applied current of $\pm4$\units{mA} we satisfy the condition that the junction is in the normal state ($I \gg I_c$), ensuring true measurement of $R_N$. $\Delta$R, presented in Figure \ref{2} (b), is the \% difference between the out-of-plane magnetic field sweeps in each direction. Due to the giant magnetoresistance effect, when the magnetic layers align antiparallel, a higher resistance state is expected compared to the parallel alignment \cite{PhysRevLett.61.2472}. We see this in the device. When the field is applied such that the magnetizations are antiparallel, the resistance of the device is about 0.25\% higher than the parallel case. Note that switching behaviour of the devices differ from the sheet films in two key ways. Firstly, the coercive fields of each layer have increased and secondly, the field range in which the device can be aligned antiparallel is much larger. It is known that in PMA thin films, switching occurs by rapid domain wall motion from nucleation sites \cite{Kikuchi2011,Metaxas2010,Vogel2006}. It is therefore to be expected that the magnetic switching of the patterned junctions occurs at higher applied fields than the sheet films due to the lower probability of a nucleation site in the small junction area.

For practicable memory devices, it will be important to reduce the low temperature switching fields of the `free' magnetic layer, and to remove the requirement for a global applied field by incorporating on-chip switching. Routes towards reducing the coercive fields of the PMA layers include refinement of the ferromagnet thicknesses and exploration of alternative materials. Routes towards on-chip switching of PMA devices include the use of on-chip coils \cite{doi:10.1021/acs.nanolett.9b02840}, which could be made from superconducting material. Spin-orbit torque switching of the barrier state could also be explored \cite{miron2010current,Liu555}.

For circular Josephson junctions, the $I_c$(B) response can be described by the Airy function \cite{barone1982physics},
\begin{equation}
\label{Airy}
I_c = I_{c0} \left | 2J_1 (\pi \Phi / \Phi_0)/(\pi \Phi / \Phi_0) \right |,
\end{equation}
\noindent where $I_{c0}$ is the maximum critical current, $J_1$ is a Bessel function of the first kind, $\Phi_0=h/2e$ is the flux quantum, and $\Phi$ is the flux through the junction, 
\begin{equation}
\label{Phi}
\Phi = \mu_0 (H_\text{app} - H_\text{shift}) w (2\lambda_\text{L} + d),
\end{equation}
\noindent where $w$, $\lambda_\text{L}$ and $d$ are the width of the junction, the London penetration depth of the electrodes\cite{London} and the total thickness of all the normal metal layers and \textit{F} layers in the junction. $H_\text{app}$ is the applied field and $H_\text{shift}$ is the amount $I_{c0}$ is shifted from $H$ = 0. $H_\text{shift}$ arises from a combination of an intrinsic contribution due to any in-plane magnetization of the junction, and extrinsic artifacts from trapped flux in the 3\units{T} superconducting coil used to perform the measurements. Fits to these equations are shown along with the data on our devices in Figure \ref{3}.

In order to study the contribution to the PSV device from each \textit{F} layer individually, we fabricated two devices each with only a single \textit{F} layer and otherwise unaltered structure from that shown in the schematic of Figure \ref{1} (a). The device with Pt(10)/CoB(0.3)/Pt(5) weak link is shown in Figure \ref{3} (a) and with Pt(5)/Co(0.6)/Pt(10) weak link in Figure \ref{3} (b). Prior to these measurements, the junctions were saturated with a 1\units{T} out-of-plane applied field at 15~K, to avoid trapping flux vorticies in the superconductor. After the field was set back to 0, the sample was rotated so the measurement field was applied in-plane and then cooled through $T_c$. The measurements were performed at the base temperature of our cryostat (1.5\units{K}).

For the single \textit{F} layer devices, Figure \ref{3} (a,b), $I_c$(B) is centered close to zero applied field as expected for \textit{F} layers with PMA. The small $H_\text{shift}$ in these data suggests there may be either trapped flux in our superconducting coils used to acquire these data or there is a small in-plane component of the magnetization. An in-plane magnetization component can arise either intrinsically or due to magnetization tilting by the applied field. This latter effect, however, cannot account for the $H_\text{shift}$, but would cause some distortion to the $I_c(B)$ at larger $H_\text{app}$. Since the single F layer junctions fit well to Eq (\ref{Airy}), this tilting effect is minimal. The CoB device is fabricated to a $3~\mu$m diameter junction and the Co to $4~\mu$m diameter, hence the tighter $I_c(B)$ pattern of the Co. The reported $I_cR_N$ is normalized to the junction area, and thus allows direct comparison despite the different junction sizes. Clearly the Co is much more transparent to supercurrent compared to the CoB, which has a much lower $I_cR_N$ despite being only half as thick. The $I_cR_N$ of the CoB appears to be the limiting factor in the $I_cR_N$ of the PSV, and hence our future research should focus on finding an alternative for the CoB \textit{F} layer. 

In Figure \ref{1} (b) we studied the influence of the $F$ layer switching on the critical current at zero applied field. In ferromagnetic Josephson junctions however, the magnetization of the \textit{F} layer can contribute to the magnetic flux density in the junction, introducing $H_\text{shift}$ proportional to the magnetization of the \textit{F} layer collinear with $H_\text{app}$. This can be significant in magnetization in-plane devices, where $\mu_0H_\text{shift}$ can be 10's\units{mT} causing the soft $F$ layer to switch prior to reaching $I_{c0}$ in the $I_c$(B) sweep \cite{PhysRevB.97.024517}. In our junctions, where the magnetization of the \textit{F} layers is perpendicular to plane, we do not expect this to occur as the layers do not contribute significant in-plane magnetization components and cannot be switched by an in-plane applied field. Nonetheless, since there is a small $H_\text{shift}$ in the junctions with single $F$ layers, it is important to measure $I_c$(B) for each state. These measurements are to confirm that the difference in critical current in Figure \ref{1} (b) is due to the PSV effect and not from any component of magnetization contributing magnetic flux density in the junction. 

In Figure \ref{3} (c-f) we study the $I_c$(B) characteristic of the device with two \textit{F} layers. To achieve the magnetic state we applied the out-of-plane fields at 15\units{K}, above $T_c$. Using the data in Figure \ref{2} (b) as a guide, a $(-)+1\units{T}$ saturating field was used for the parallel magnetic alignments. For the antiparallel alignments, the $(-)+1\units{T}$ saturating field was followed by a set field of $(+)-0.3\units{T}$. As per the single $F$ layer samples, the field was set back to 0, the sample was rotated, so that the measurement field was applied in-plane, and then cooled to 1.5\units{K}. 

The measurements of $I_c$(B) confirm that in the parallel alignment the $I_{c0}$ is significantly smaller than in the antiparallel alignment. The maximum value of $I_{c} R_N$ measured from the two parallel states was $0.312\pm 0.001~\mu$V. By contrast, in the antiparallel state the maximum $I_{c}R_N$ measured was $0.504\pm 0.001\units{\mu V}$. This corresponds to a $\Delta I_c/I^{para}_c = 60.7\pm0.2\%$. In a memory scheme, the high/low $I_c$ states of the device form a memory bit. By measuring at a constant applied current larger than $I_c$ parallel and less than $I_c$ antiparallel, the device will either be in the superconducting or resistive state. It is also possible that the two states (parallel and antiparallel) correspond to different ground-state phase differences across the junction (0 or $\pi$), although our measurements here are not sensitive to such a change. We have summarized our experimental results in Table \ref{table1}.

In conclusion, we report on pseudospin-valve Josephson junctions with perpendicular magnetic anistropy. The Co layer with $0.6\units{nm}$ thickness and CoB layer with $0.3\units{nm}$ thickness have sufficiently independent switching to allow the device to be placed in either the parallel or antiparallel magnetic configuration. The critical current of the device is strongly dependent on the magnetic configuration, with a (high)low critical current state in the (anti)parallel configuration. This demonstration shows that our work has immediate device application as a memory bit in an all superconducting computation scheme. In the future it will be possible to map a phase diagram for PSV Josephson junctions with perpendicular magnetic anisotropy to allow control of the ground-state phase differences between 0 or $\pi$, and explore the use of spin-orbit coupling for additional device functionality.

The data associated with this paper are openly available
from the University of Leeds data repository \cite{NoteY}.

\begin{acknowledgments}
We wish to thank Norman Birge and Reza Loloee for advice and helpful discussions, T. Moorsom, A.J. Huxtable, A. Walton and M. Rogers for experimental assistance. We acknowledge support from the Henry Royce Institute. The work was supported financially through the following EPSRC grants: EP/M000923/1, EP/P022464/1 and EP/R00661X/1. M Algarni acknowledges support from Jeddah University. This project has received funding from the European Unions Horizon 2020 research and innovation programme under the Marie Sk\l{}odowska-Curie Grant Agreement No. 743791 (SUPERSPIN).

\end{acknowledgments}

\bibliography{library}

\end{document}